\documentclass{iopconfser}
\usepackage{float}
\usepackage{graphicx}
\usepackage{subcaption}
\usepackage[backend=biber,sorting=none,sortcites=true,citestyle=numeric-comp]{biblatex}
\begin{document}

\title{Transient CFD simulation of a linear start-up of a variable speed reversible pump-turbine}

\author{Aron D. Beck$^{1*}$, Alexander Masero$^{2}$, Ernesto Casartelli$^{2}$ and Elena Vagnoni$^{1}$}

\affil{$^1$Technology Platform for Hydraulic Machines (PTMH), EPFL, Lausanne, Switzerland}
\affil{$^2$Competence Centre of Fluid Mechanics and Numerical Methods, HSLU, Horw, Switzerland}

\email{aron.beck@epfl.ch$^{*}$}

\begin{abstract}

The integration of intermittent energy sources has led to a growing demand for fast response and partial-load operation of hydraulic turbines, making variable-speed units increasingly important. In hydraulic turbines, frequent start-stop cycles generate the highest loads under normal operating conditions. Although historically infrequent and often neglected in turbine design, the growing occurrence of these cycles now necessitates particular consideration due to their contribution to fatigue loading and reduced runner lifetime. Experimental investigations of turbine start-up are costly and cannot be conducted during early design stages. While start-up sequences have been simulated with CFD for fixed-speed turbines, each approach has had notable limitations. Some studies have lacked experimental validation. Some studies have used reduced geometrical domains, making it unclear whether inaccuracies stem from geometry simplification or turbulence modelling, and others employ hybrid LES/RANS models, at a very high computational cost, making them unfeasible for use during the design phase. In this study, the linear start-up of a variable-speed turbine start-up is simulated using URANS-based approaches. Simulations are performed using the k-$\omega$ SST, EARSM, and the RSM turbulence models within a computationally tractable framework. The simulations show that torque, mass flow, and the blade-passing frequency content at the leading edge are accurately captured by all models. In contrast, the frequency content at the trailing edge and in the draft tube is considerably better resolved using the RSM than the other models because of its capability to consistently capture turbulence anisotropy. 

\end{abstract}


\section{Introduction}

Due to the growing use of hydropower plants for grid regulation and the corresponding increase in start/stop cycles \cite{alerci_modeling_2025}, the focus on methods for fatigue estimation is increasingly important. Computational Fluid Dynamics (CFD) has been the main tool for performance and load estimation during the design phase for hydraulic turbomachines and is used both in steady state and transient formulations. It has, though, mainly been used at steady operation of the machine for performance estimation due to high computational cost associated with simulations featuring hydraulic transients such as start-stop operations or load rejections. However, in recent years, with the development of faster and more powerful computers, CFD studies of transient operations are gaining popularity for analysis of operational issues and instabilities that might arise during load rejections or turbine shut-downs~\cite{mossinger_transient_2016,mossinger_francis-99_2017}. For start-up sequences, CFD studies have been conducted for fixed-speed turbines. However, each approach has notable limitations. Some studies lack experimental validation where pressure pulsations are considered \cite{casartelli_cfd_2019,sun_numerical_2024}. Others use simplified geometries, making it difficult to determine whether discrepancies with experiments arise from the geometric simplifications or the turbulence modelling approach \cite{morissette_transient_2012,morissette_fluid-structure_2019}. Finally, several studies employ hybrid LES/RANS turbulence models, which capture the flow physics with high accuracy but at a very high computational cost, exceeding 150,000 core-hours \cite{zorn_cfd_2024,salehi_flow-induced_2022,morissette_fluid-structure_2019,unterluggauer_experimental_2020}. Therefore, the feasibility of URANS turbulence models featuring the full geometry remains uncertain, as well as there are no studies that consider the fully transient approach for a start-up in a variable-speed turbine.  


In this paper, we present a comparison of unsteady simulations performed with the block-coupled solver coupledNumerics~\cite{mangani_development_2014} using the k-$\omega$ SST, EARSM and RSM turbulence models for a linear start-up in a variable-speed, reduced-scale pump-turbine and compare to experimental data acquired for the corresponding sequence in transient at one of PTMH's platforms featuring discharge, torque and pressure fluctuations on the runner and in the draft tube. To make the comparison as representative as possible, boundary conditions are applied in transient to take into account the up- and downstream pressure fluctuations during the start-up. Additionally, a dynamic mesh is used to impose the opening of the guide vanes, which employs mesh deformation instead of remeshing at each timestep to speed up computations. 

\section{Methodology}

\subsection{Numerical set-up}
\begin{figure}[H]
\begin{minipage}{0.5\linewidth}
    \centering
    \includegraphics[width=1\textwidth]{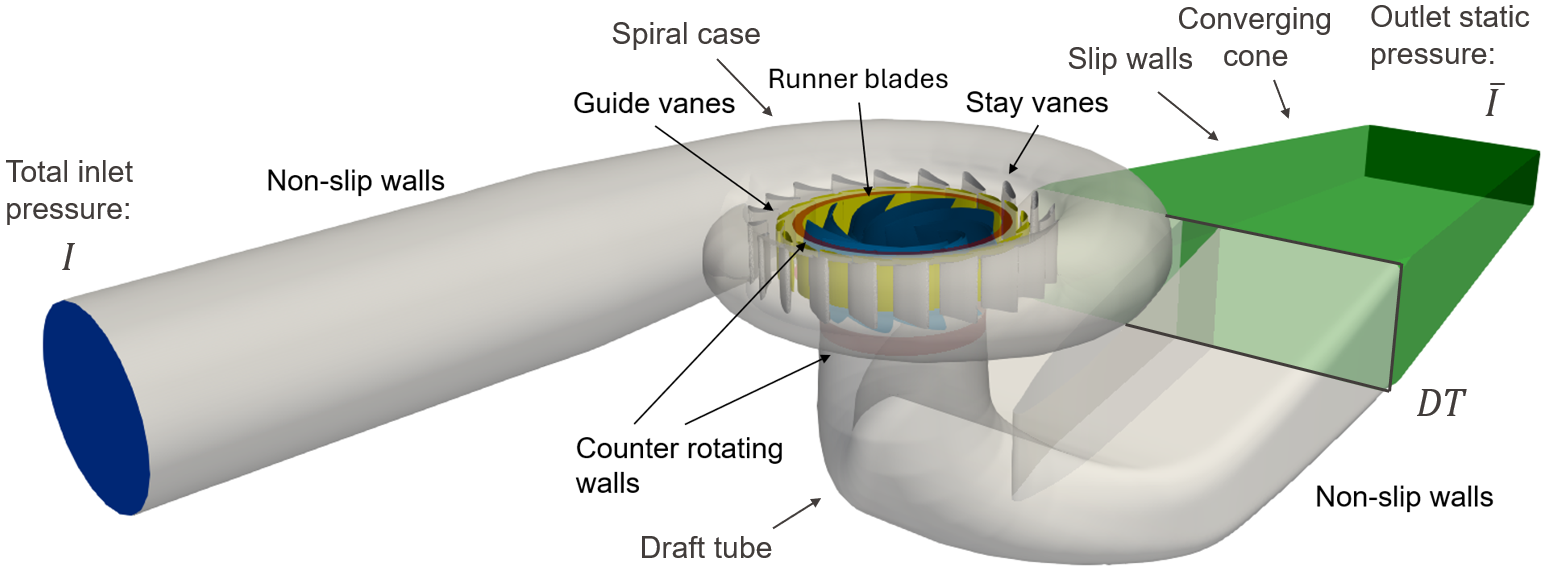}
    \subcaption{\label{fig:bc-domain}Simulation domain}
    \end{minipage}%
    \hfill
    \begin{minipage}{0.5\linewidth}
    \centering
    \includegraphics[width=1\textwidth]{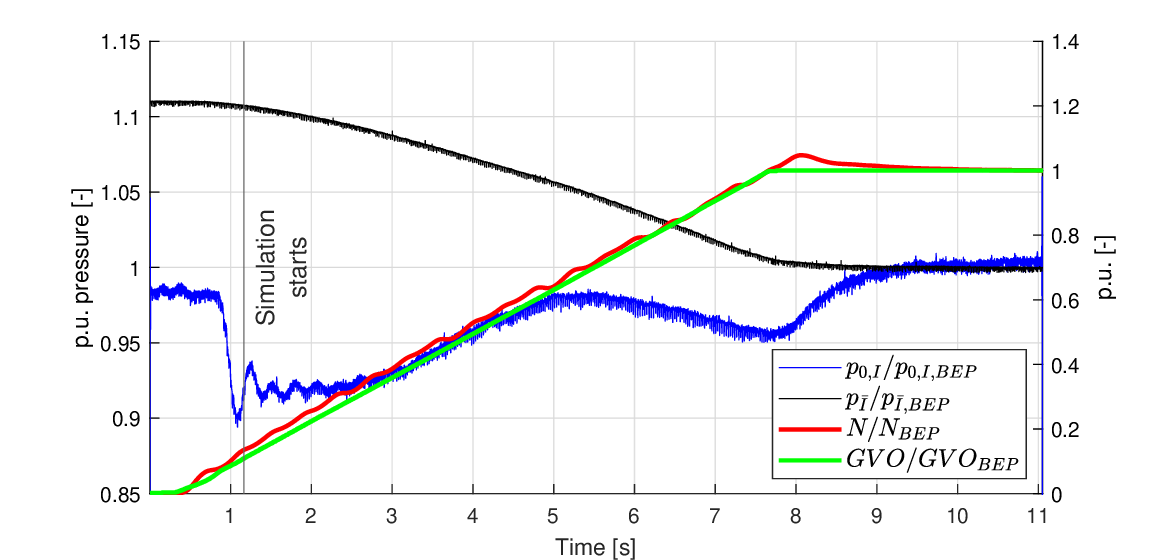}
    \subcaption{\label{fig:bc-transient}Transient boundary conditions}
\end{minipage}%
\caption{\label{fig:numerical-setup}Simulation set-up}
\end{figure}

The simulation domain in this study considers a model-scale reversible pump-turbine which features 9 blades and 20 guide vanes and can be appreciated in Figure \ref{fig:bc-domain}. The full geometry includes the spiral case, stay and guide vanes, the impeller, and the draft-tube. To avoid backflow from the outlet, a converging section (green transparent in Figure \ref{fig:bc-domain}) was added to the fluid domain to accelerate the fluid. 

The  boundary conditions can be appreciated in Table \ref{tbl:bc-fluid}. For the simulation, the total pressure is imposed at the inlet, $I$ and the corrected static pressure is used at the outlet $\bar{I}$, by applying the Bernoulli principle (eq.\ref{eq:pout}). At the walls, no-slip velocity is imposed. As it is important to have elements of high orthogonality at the sliding interfaces where the fluid information has to be transferred between the rotating domain and the guide vane domain, the boundaries of the rotating domain were moved slightly into the guide vane domain. This ensured that the boundary was farther from the blades and that the element quality at the interface would be better.
To counterbalance the movement of the rotating domain in those extension regions, counter-rotating velocity was imposed at the walls so that they have zero velocity in the absolute reference frame. These regions are indicated with a transparent red colour in Figure \ref{fig:bc-domain}.  

\begin{table}[H]
    \centering
        \caption{Boundary conditions}
        \footnotesize
    \label{tbl:bc-fluid}
    \begin{tabular}{|c|c|c|c}\hline
    Boundary & Colour in Fig. \ref{fig:bc-domain} &Type  \\ \hline \hline
    Inlet & Dark blue & Total pressure \\ 
    Outlet & Green & Static pressure \\
    Runner & Light blue & No-slip walls  \\
    Guide vanes & Yellow & No-slip walls  \\
    Stay vanes & White & No-slip walls \\
    Spiral case & White transparent    & No-slip walls  \\
    Draft tube & White transparent     & No-slip walls \\
    Counter rotating patches& Red transparent & Moving wall \\
    Outlet converging section & Green transparent & Slip walls  \\ \hline
    \end{tabular}
\end{table}

The static outlet pressure is defined by Equation \ref{eq:pout}, taking into account the pressure drop in the converging section at the outlet, which was added to reduce backflow in the simulation. Slip walls were added to this zone to minimize its effect on the results of the simulation.

\begin{equation}
    p_{\bar{I}} = \Delta p_\sigma + p_{atm} - \Delta p_{c}
    \label{eq:pout}
\end{equation}

\begin{figure}[H]
    \centering
    \includegraphics[width=0.5\linewidth]{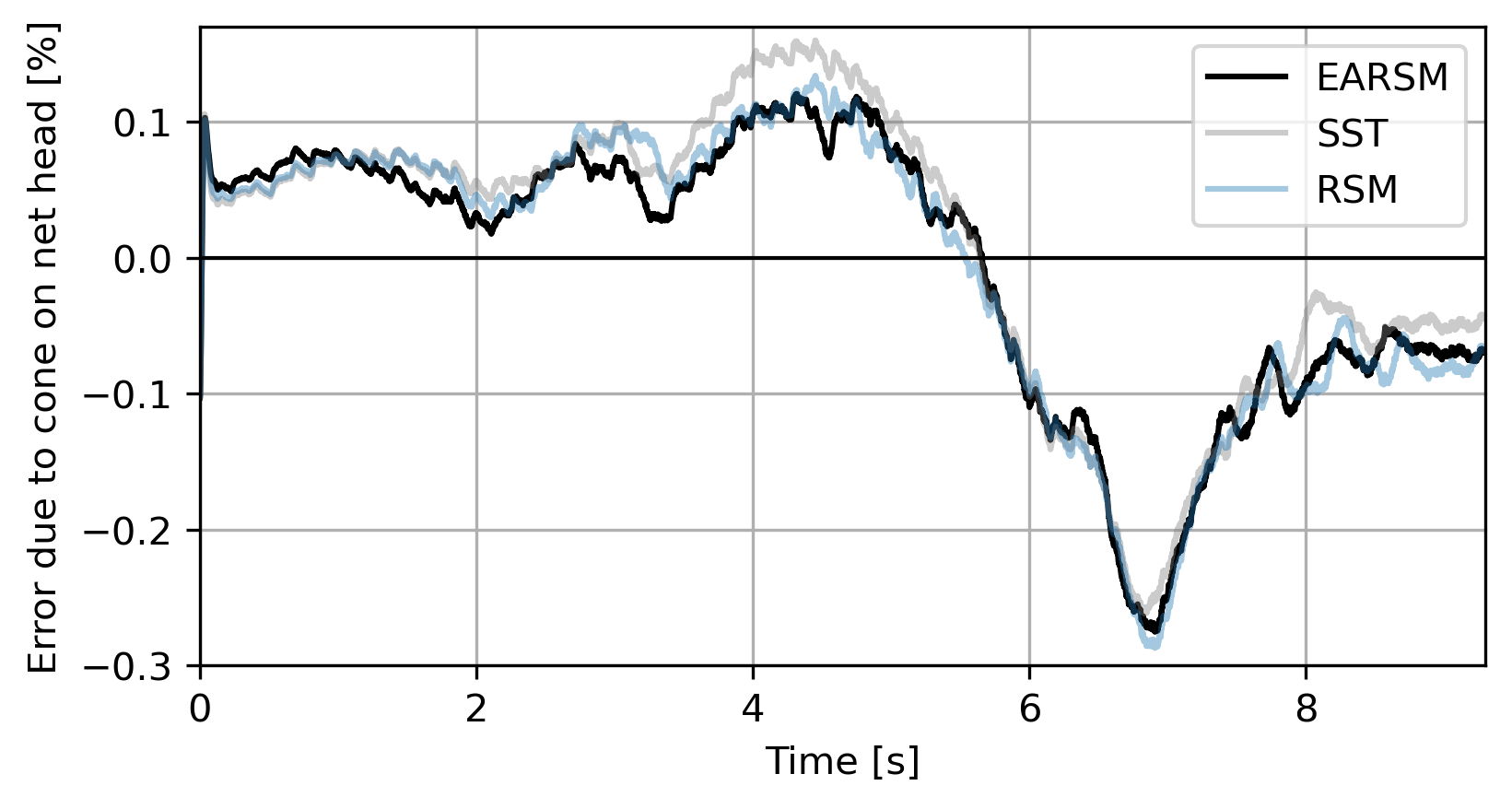}
    \caption{Error due to cone on net head, ($H_{sim}$ - $H_{exp}$)/$H_{exp}$)*100 \%}
    \label{fig:error-cone}
\end{figure}

In equation \ref{eq:pout}, $\Delta p_\sigma$ is the suction pressure, $p_{atm}$ the atmospheric pressure, and $\Delta p_{c}$ the inviscid and incompressible pressure drop in the outlet converging section derived from the Bernoulli equation. The suction and the atmospheric pressures were measured during the experiment, while the pressure correction term can be derived as follows:

\begin{equation}
    \Delta p_{c} = p_{r,o,s} - p_{r,o,DT} = \frac{1}{2} \rho Q^2 \left(\frac{1}{A_{o,s}^2}-\frac{1}{A_{o,DT}^2}\right)
    \label{eq:bernoulli}
\end{equation}

where $p_{r,o,s}$ is the relative outlet pressure of the simulation domain, $p_{r,o,DT}$ the relative pressure at the draft tube outlet (before the converging section), and $\rho$ the density of the water. The discharge $Q$, was then used to define the pressure drop based on the sectional area of the simulation outlet $A_{o,s}$, and the sectional area of the draft tube $A_{o,DT}$. By using slip walls and the pressure correction in equation \ref{eq:bernoulli}, the overestimation of the head caused by the converging outlet section at best efficiency operation (BEP) was minimized to a value of 0.08\%, which can be associated with turbulence generation that the CFD models solve for but the Bernoulli equation neglects. At the inlet section, the total pressure is defined as follows:
\begin{equation}
    p_{0,I} = \Delta p_{H,M} + \Delta p_{\sigma} + p_{atm}+ \frac{1}{2}\rho\frac{Q^2}{A_{DT}^2}
\end{equation}
where $p_{H,M}$ is the head pressure measured from the experiment, between the spiral case inlet and the draft tube outlet section. The outlet kinetic energy is added to correct the head between the inlet and draft tube outlet. The error introduced by adding the converging section to the simulation domain while considering inlet total pressure with an outlet static pressure is negligible and can be appreciated in Figure \ref{fig:error-cone}. 

The fluid domain, apart from the guide vane region, was meshed using the Poly-Hexcore (P-H)  approach available in Fluent meshing \cite{ansys_fluent_2024}. The boundary layer of the unstructured mesh had an expansion ratio of 1.2, and all walls had 8 inflation layers. A structured hex mesh was used in the guide vane domain due to the guide vane movement. The guide vane domain was meshed using TurboGrid \cite{ansys_turbogrid_2024}. The fluid domain mesh had a total element count of 9.3M elements.


\begin{table}[H]
    \centering
    \scriptsize
    \caption{Mesh quality report}
    \label{tab:mesh-quality}
    \begin{tabular}{|c|c|c|c|c|c|c|c|c|} \hline
        Cell Zone & Sim Type & El. type & El. count & Min ortho. & Avg. ortho. & Max skew. & Avg. skew & y$^{+ ^*}$\\ \hline \hline
        Runner & Fluid & P-H & 4,945,772 & 0.22 & 0.92 & 0.73 & 0.06 & 52.26\\ 
        Guide vanes & Fluid & Hex & 708,560 & 0.88 & 0.99 & 0.37 & 0.05 & 39.2 \\ 
        Spiral case & Fluid & P-H & 2,209,033 & 0.22 & 0.93 & 0.77 & 0.05 & 111.52 \\
        Draft tube & Fluid & P-H & 1,326,077 & 0.32 & 0.98 & 0.72 & 0.02 & 120.3\\ \hline
    \end{tabular}\\[1ex]
    \raggedright
    *At BEP
\end{table}

\begin{figure}[H]
\centering
\begin{minipage}{0.4\linewidth}
    \centering
    \includegraphics[width=0.9\textwidth]{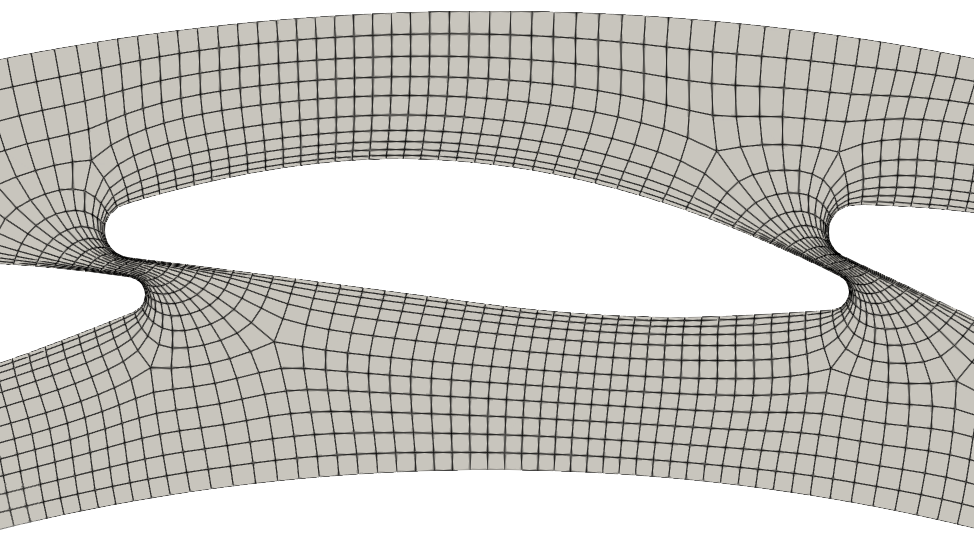}
    \subcaption{\label{fig-1_6}Guide vane domain at 1.6$^{\circ}$}
    \end{minipage}%
    \begin{minipage}{0.4\linewidth}
    \centering
    \includegraphics[width=0.91\textwidth]{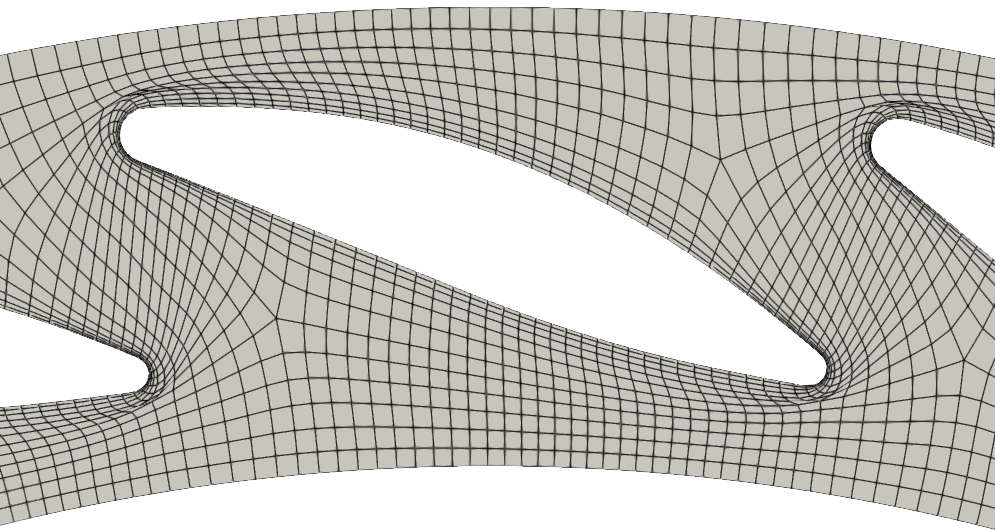}
    \subcaption{\label{fig-14_5}Guide vane domain at 14.5$^{\circ}$}
\end{minipage}%
\caption{\label{fig:mesh-fluid}Mesh deformation at extreme openings}
\end{figure}

To impose the rotation of the runner, a sliding mesh approach is used. Since the guide vanes will open during the simulation, a dynamic deforming mesh is used. In the simulation, the deformation is performed from 1.6$^{\circ}$ to 14.5$^{\circ}$. The mesh at those extreme openings can be seen in Figures \ref{fig-1_6} and \ref{fig-14_5}. The value of 1.6$^{\circ}$ was selected as the opening nearest to being completely closed without elements collapsing together. To allow the mesh to deform between those extreme openings, the guide vane domain was meshed at 5$^{\circ}$ guide vane opening. In a first attempt, the guide vane domain was meshed at half of the opening range during the start-up, but a sensitivity study revealed that an opening of 5$^{\circ}$ was optimal as a starting point in order to close the guide vanes as much as possible and being able to deform into BEP position while maintaining adequate mesh quality.

Three different turbulence models are considered and used for comparison. The $k-\omega$ SST was considered due to its widespread use and good accuracy reported at BEP \cite{deniz_experimental_2019}. The BSL-EARSM was additionally considered since it was reported as having the best accuracy of two-equation RANS turbulence models at speed-no-load (SNL) when compared to experimental results ~\cite{deniz_experimental_2019}. Due to high swirl expected in the draft tube, the full baseline Reynolds stress model (BSL-RSM) is additionally considered, which is capable of solving for the individual components of the Reynolds stress tensor but requires 5 additional transport equations \cite{ansys_fluent_fluent_2025}.

 All models are solved in transient and with a second-order spatial and time discretization. For all models, a wall function is used to model the boundary layer, as commonly done in CFD simulations of hydraulic machines since simulations become significantly heavier when resolving the flow of the boundary layer. For the mesh, it was targeted to keep $y^+>$ 30 in the log-law region and avoid the buffer layer \cite{ansys_ansys_2009}.  In the simulations, a time step equivalent to 4$^{\circ}$ of rotation at BEP was selected. The simulation of the start-up sequence was run using 5x Intel Xeon E5-2630 v4 on a total of 90 cores and took 16 days to solve for the two-equation models. That is the equivalent of 34.560 core hours. The BSL-RSM turbulence model required 25 days or 56.160 core hours.

\subsection{Experimental set-up}

\begin{figure}[H]
\centering
\begin{minipage}[t]{0.62\linewidth}
    \centering
    \includegraphics[width=0.8\textwidth]{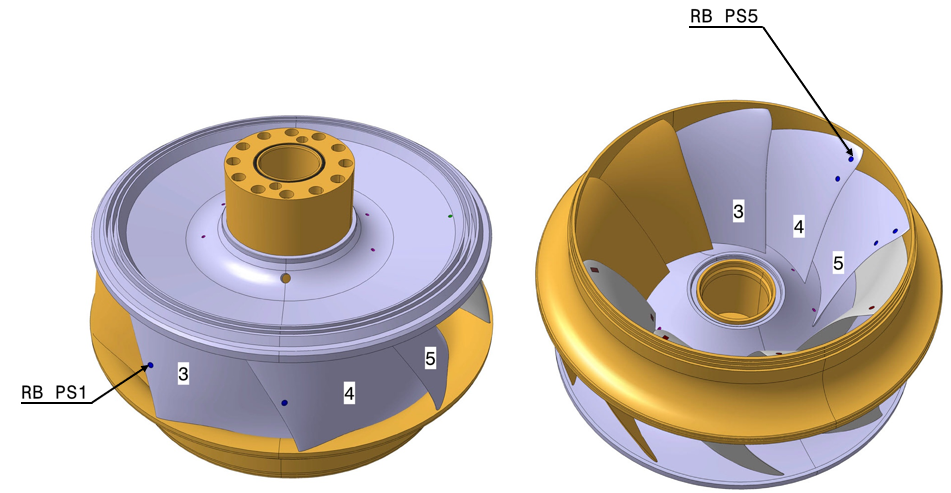}
    \subcaption{\label{fig-stress_sensors}Location of on-board pressure sensors}
\end{minipage}%
\hfill
\begin{minipage}[t]{0.37\linewidth}
    \centering
    \includegraphics[width=0.75\textwidth]{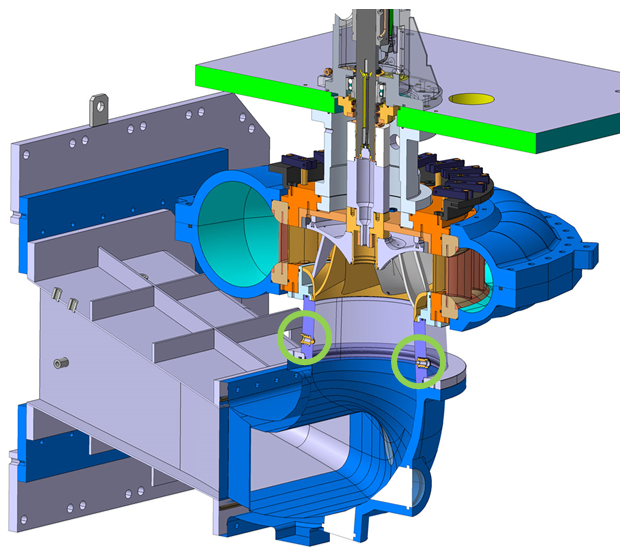}
    \subcaption{\label{fig-pressure_sensors}Location of static pressure sensors}
\end{minipage}%
\caption{\label{fig-exp} Position of the pressure sensors in the reduced scale model\cite{seydoux_assessments_2022}}
\end{figure}

The results of the reduced-scale model experiments conducted by Seydoux et al. \cite{seydoux_experimental_2024} are used to validate the CFD model. The experiment features a transient measurement of the linear start-up sequence, which was conducted at one of the test-rigs at the EPFL Technology Platform for Hydraulic Machines in accordance with IEC-60193 using a closed-loop configuration with pumps and a variable speed generator capable of simulating the realistic response of the hydraulic circuit at the power plant. The pressure sensors used for comparison are the pressure transducers (UNISENSOR-HF900-03291) in the cone above the draft tube (Figure \ref{fig-pressure_sensors}) and on the runner (RB PS1 and RB PS5) as shown in Figure \ref{fig-stress_sensors}. Both have an acquisition frequency of 1000 Hz and an uncertainty of 0.7\%. 


\subsection{Post-processing of CFD - Dimensionless factors}

To analyse the model performance and simulation results,  the following dimensionless IEC factors \cite{iec_iec_nodate} are used, namely the discharge, $Q_{ED}$ rotational speed, $n_{ED}$ and the torque, $T_{ED}$: 
\begin{equation}
    Q_{ED} = \frac{Q}{D^2\sqrt{E}}
\end{equation}

\begin{equation}
    n_{ED} = \frac{nD}{\sqrt{E}}
\end{equation}

\begin{equation}
    T_{ED} = \frac{T_m}{\rho D^3E}
\end{equation}

where $D$ is the runner's inlet diameter, $\rho$ the water's density, and $E = gH$ the specific total energy, with $H$ the net head measured in the platform, as well as the discharge, $Q$ and the mechanical torque, $T_m$. 
Additionally, the pressure coefficient is computed as follows:

\begin{equation}
    c_p = \frac{p-p_{ref}}{\rho E}
    \label{eq:cp}
\end{equation}
with $p$, the static pressure and $p_{ref}$ the reference pressure. In this study, it will be the average pressure.

\section{Results}
\subsection{Time domain comparison to measurements}
\begin{figure}[H]
    \centering
    \includegraphics[width=1\linewidth]{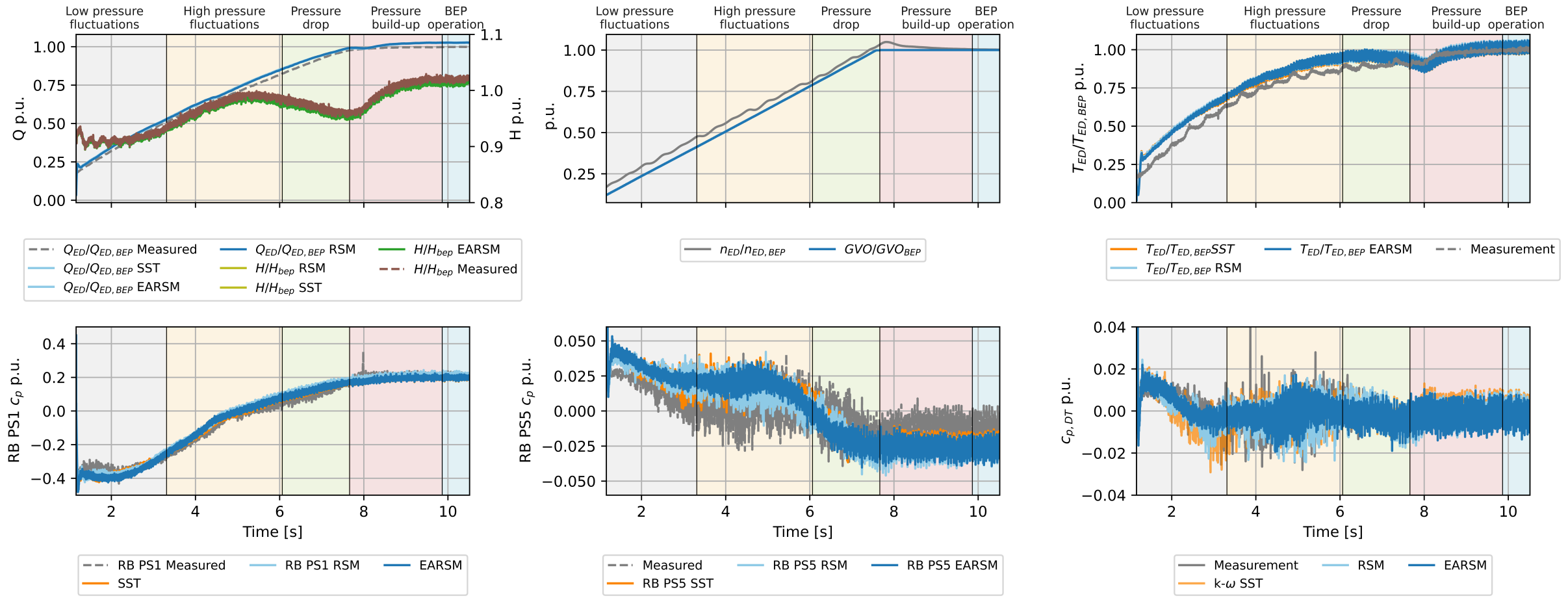}
    \caption{Comparison between probe measurements showing discharge and head variation in the top-left subfigure, the rotational speed and the guide vane opening in the top-middle and the torque in the top-right. The leading edge pressure coefficient is shown in the bottom-left figure, the trailing edge pressure coefficient in the bottom-middle and the draft tube pressure coefficient in the bottom-right.}
    \label{fig:start-up-platform}
\end{figure}

A comparison with the platform measurements considering the measured torque $T_{ED}$, discharge $Q_{ED}$, and the pressure coefficient on the draft tube walls $c_{p,DT}$, shown in Figure \ref{fig-pressure_sensors}, were conducted to validate the model performance in simulating the linear start-up sequence. The results are shown in Figure \ref{fig:start-up-platform}. Five main flow regimes identified along the trajectory from the pressure measurements in the draft tube will be used to analyse the flow and validate the CFD simulations. Those five main regimes are marked by shaded colours in Figure \ref{fig:start-up-platform}. The following regimes are: 
\begin{itemize}
    \item \textbf{Low-pressure fluctuation} regime is present between 0 - 3.5 s. 
    \item \textbf{High-pressure fluctuation} regime is between 3.5 - 6 s, high instability is observed in the draft tube due to vortices.
    \item \textbf{Pressure drop} regime is between 6 - 7.5 s. This regime is characterised by the collapse of the draft tube vortices into uniform flow.
   \item \textbf{Pressure build-up} is between 7.5 - 9.7 s, the pressure in the draft tube increases to reach BEP pressure where the machine operating condition stabilizes to the set head and discharge.
    \item \textbf{BEP} operating point is reached at 9.7 s. This is characterized by a stable flow condition and low-amplitude pressure fluctuation. 
\end{itemize}

 The evolution of torque and mass flow, illustrated in Figure \ref{fig:start-up-platform}, demonstrates that the numerical simulations stabilize rapidly after being initialized with partially open guide vanes, confirming that the use of mesh deformation and initialization shortly after guide vane opening is valid. The mass flow evolution from the simulations follows the same trend as the measured value, but with a 3.2\% error on average between the signals is observed. The simulated torque also follows the measured torque closely with an average error of 5\% compared to the measured values but rotor side spaces were neglected in the domain, which might partially explain the deviation. However, this behaviour is in good agreement with other RANS CFD studies reported in the literature, where the models additionally struggle to capture surface roughness and turbulence effects, leading to an overprediction of torque (\cite{bergmann-paulsen_fsi-analysis_2012,trivedi_effect_2013}). Although the simulated torque reproduces the overall trend well, it does not capture the small drops observed in the measured signal, but around T = 8 s, the simulated torque manages to capture the drop in the torque before reaching BEP operation.

 The pressure coefficient at the leading edge RB PS1 and the trailing edge RB PS5 of the runner were additionally compared with measurements from absolute pressure transducers mounted on the runner, as shown in Figure \ref{fig:start-up-platform}. At the leading edge, the simulation captures the same trends and fluctuations as the measured signal, with a similar amplitude. This indicates a strong capability of the model to reproduce rotor-stator interaction (RSI) throughout the entire start-up sequence. At the trailing edge, the models perform less accurately, but they show a slightly higher pressure drop of approximately 25\% than observed in the measurements. The amplitude is also noticeably lower for the CFD models, especially for the BSL-EARSM turbulence model when the turbine enters the high-pressure fluctuation regime. The k-$\omega$ SST and the BSL-RSM perform better, although they show slightly lower amplitude compared to the measurements. Similar difficulties have been reported in the literature, where CFD simulations using more advanced SAS turbulence models also have had limited performance in accurately reproducing pressure and strain oscillations at the trailing edge of runner blades at partial load under resonance (\cite{biner_fatigue_2024, morissette_fluid-structure_2019}). Further investigation is required to determine whether these discrepancies arise from limited turbulence resolution, excessively large time steps in the simulation, or high-frequency resonance along the blade, as suggested by Biner et al. 2024 \cite{biner_fatigue_2024}. 

 \vspace{-10pt}
\vspace{-10pt}

A comparison between the simulated and measured pressure signal $c_{p,DT}$, in the draft tube provides insight into how well the models capture the instabilities caused by the draft tube vortices. During the low-amplitude regime at early stages of the start-up, the BSL-EARSM model follows the measured signal most closely, while the BSL-RSM exhibits slight numerical oscillations and a slight premature pressure drop. The k-$\omega$ SST model also exhibits these numerical oscillations and the premature pressure drop, but they appear more severe than with the BSL-RSM. However, when the turbine goes through the high-amplitude pressure pulsation regime, the BSL-RSM and the k-$\omega$ SST simulations are able to match the amplitude of the measured signal more closely than the BSL-EARSM simulation, which shows a slightly lower amplitude. All turbulence models are able to capture the trend of the pressure drop and the pressure build-up to BEP, although they show slightly higher amplitude at BEP compared to the measurement data. 

\subsection{Frequency domain analysis}
\begin{figure}[]
\centering

\begin{minipage}{0.33\linewidth}
        \centering
        \includegraphics[width=1\textwidth]{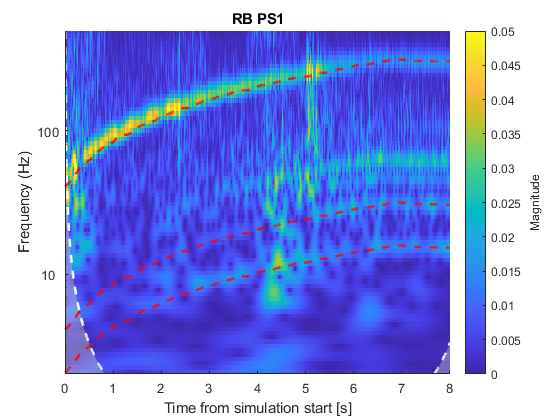}
        \subcaption{\label{fig:RB_PS1EXP}RB PS1 - Experiment}
    \end{minipage}%
    \hfill
        \begin{minipage}{0.33\linewidth}
        \centering
        \includegraphics[width=1\textwidth]{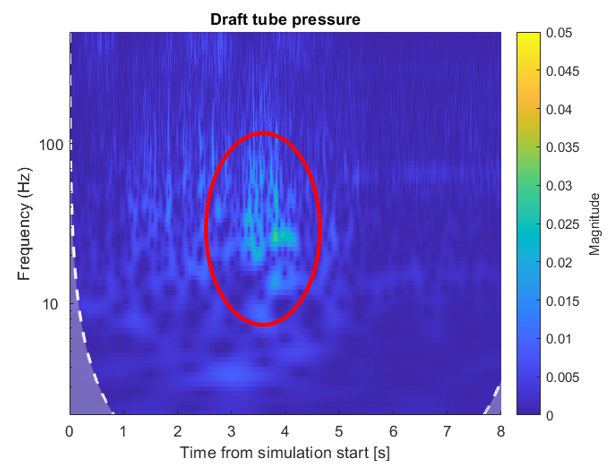}
        \subcaption{\label{fig:p9356_EXP}Draft tube - Experiment}
    \end{minipage}%
    \hfill
        \begin{minipage}{0.33\linewidth}
        \centering
        \includegraphics[width=1\textwidth]{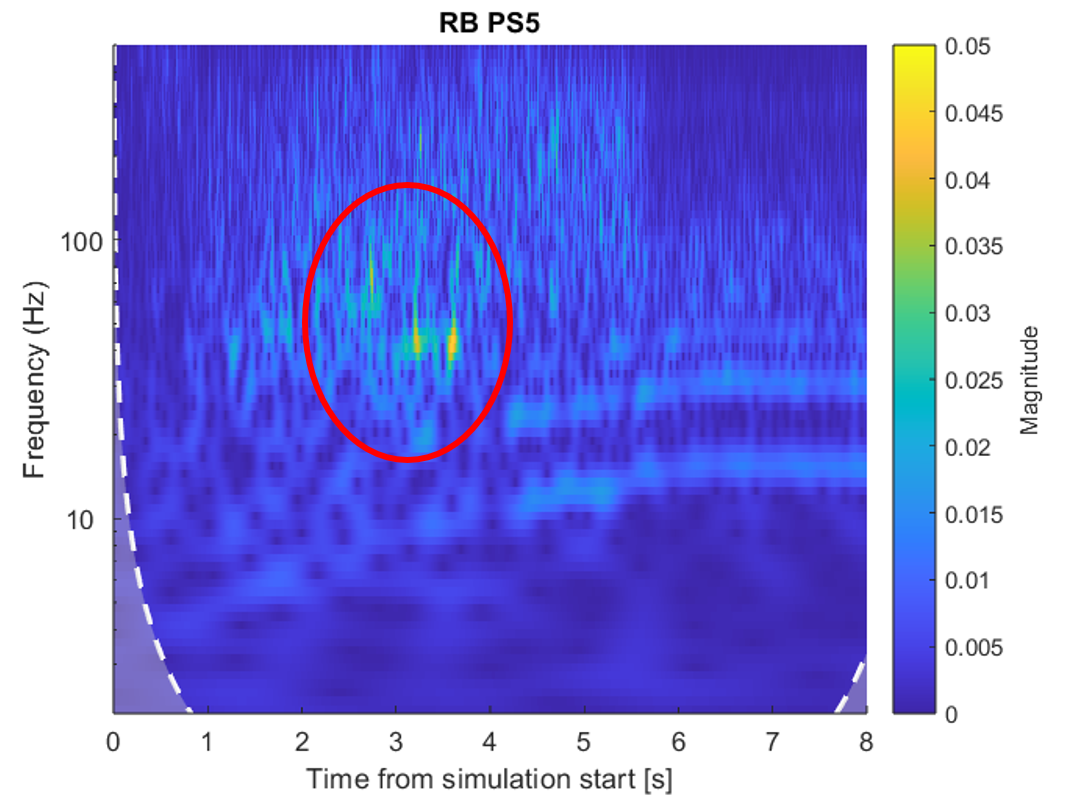}
        \subcaption{\label{fig:RB_PS5EXP}RB PS5 - Experiment}
    \end{minipage}%
    
      \hfill
    \begin{minipage}{0.33\linewidth}
        \centering
        \includegraphics[width=1\textwidth]{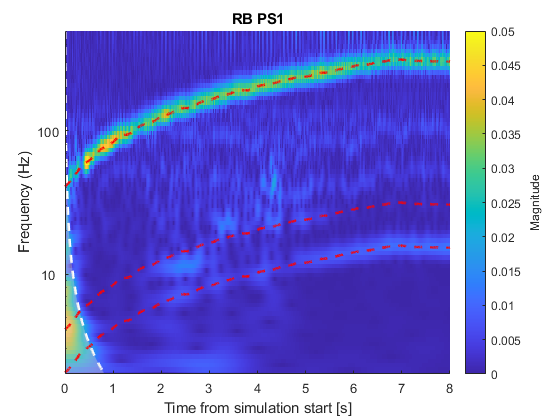}
        \subcaption{\label{fig:RB_PS1SST}RB PS1 - SST}
    \end{minipage}
       \hfill
        \begin{minipage}{0.33\linewidth}
        \centering
        \includegraphics[width=0.98\textwidth]{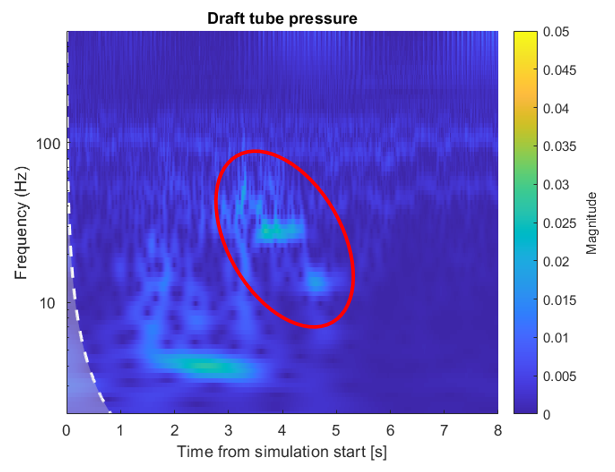}
        \subcaption{\label{fig:SST_p9356_SIM}Draft tube - SST}
    \end{minipage}%
        \hfill
        \begin{minipage}{0.33\linewidth}
        \centering
        \includegraphics[width=0.98\textwidth]{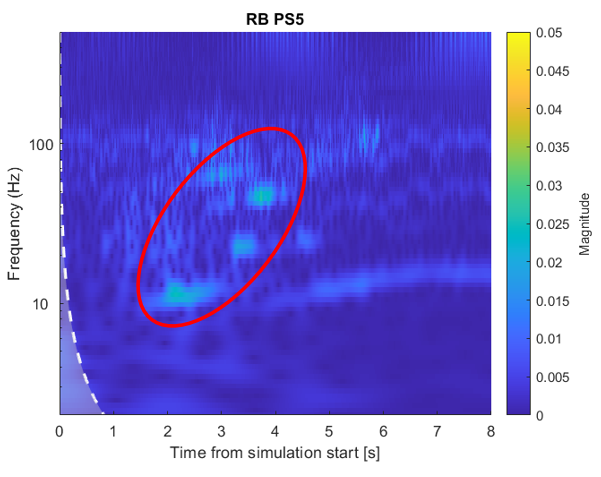}
        \subcaption{\label{fig:EARSM_RB_PS5SIM}RB PS5 - SST}
    \end{minipage}%
    \hfill
    \begin{minipage}{0.33\linewidth}
        \centering
        \includegraphics[width=1\textwidth]{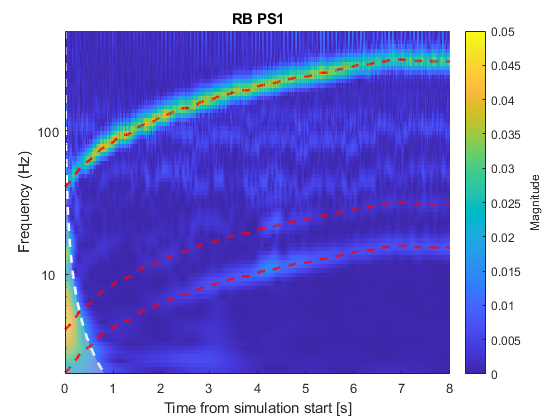}
        \subcaption{\label{fig:RB_PS1SIM}RB PS1 - EARSM}
    \end{minipage}
       \hfill
        \begin{minipage}{0.33\linewidth}
        \centering
        \includegraphics[width=0.98\textwidth]{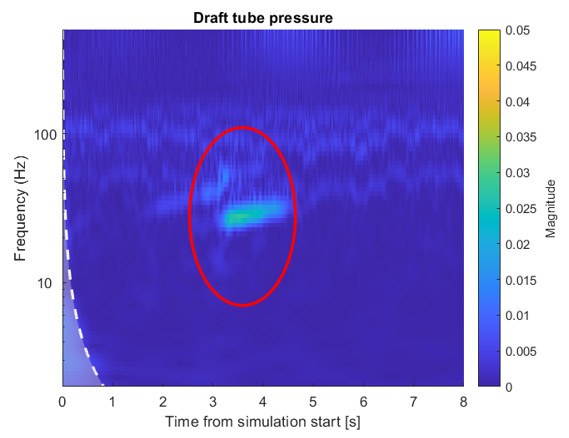}
        \subcaption{\label{fig:p9356_SIM}Draft tube - EARSM}
    \end{minipage}%
        \hfill
        \begin{minipage}{0.33\linewidth}
        \centering
        \includegraphics[width=0.98\textwidth]{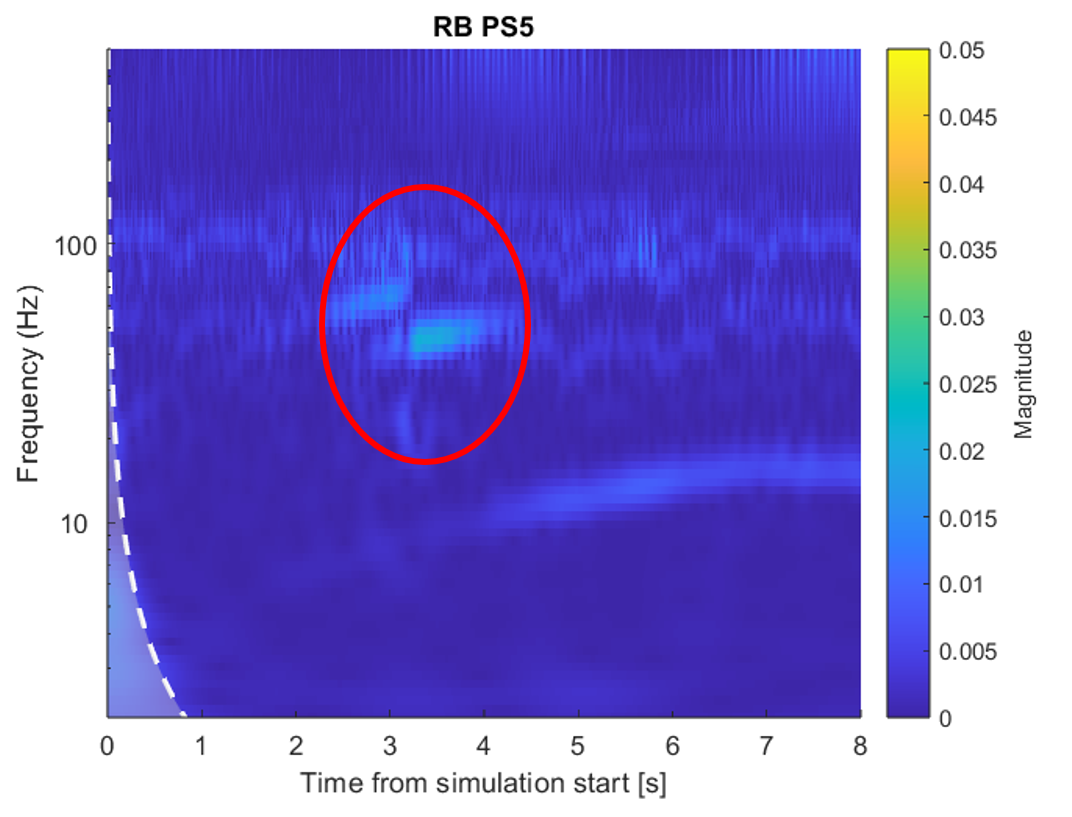}
        \subcaption{\label{fig:EARSM_RB_PS5SIM}RB PS5 - EARSM}
    \end{minipage}%
            \hfill
        \begin{minipage}{0.33\linewidth}
        \centering
        \includegraphics[width=0.98\textwidth]{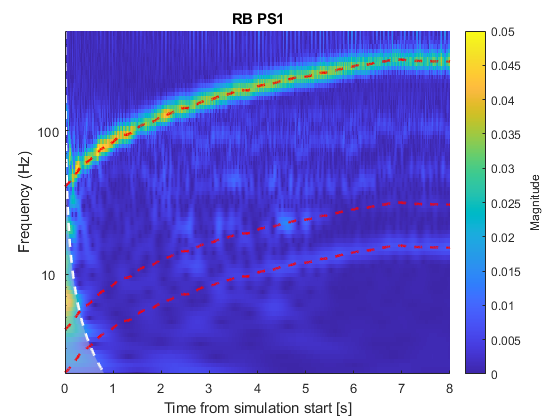}
        \subcaption{\label{fig:RB_PS1RSM}RB PS1 - RSM}
    \end{minipage}%
                \hfill
        \begin{minipage}{0.33\linewidth}
        \centering
        \includegraphics[width=0.98\textwidth]{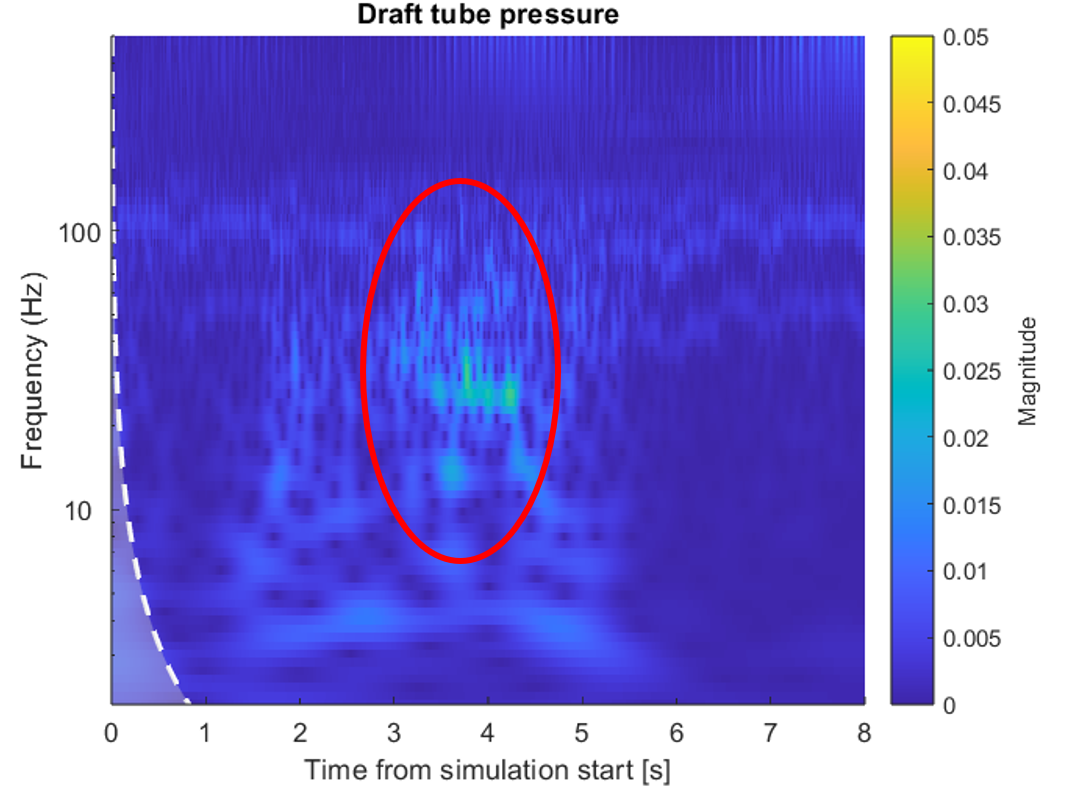}
        \subcaption{\label{fig:RSM_p9356SIM}Draft tube - RSM}
    \end{minipage}%
                \hfill
        \begin{minipage}{0.33\linewidth}
        \centering
        \includegraphics[width=0.98\textwidth]{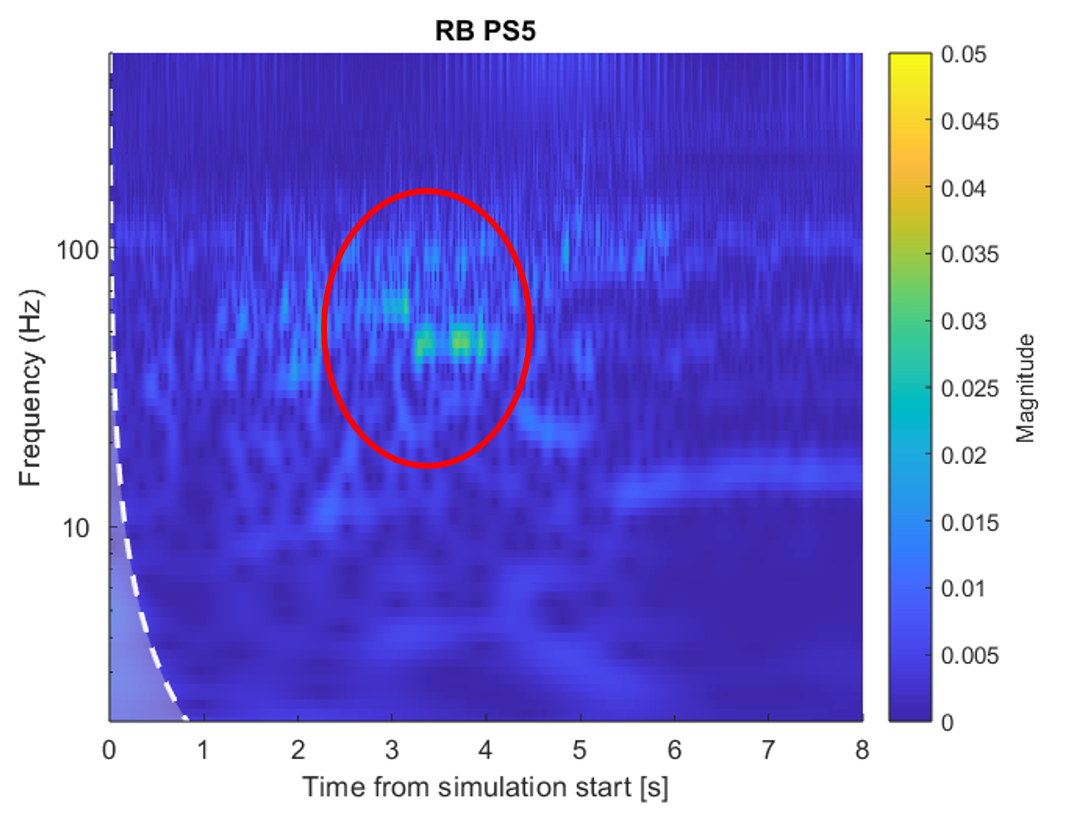}
        \subcaption{\label{fig:RSM_RB_PS5SIM}RB PS5 - RSM}
    \end{minipage}%

\caption{\label{fig:CWT-runner}Frequency comparison  with CWT scalogram of simulation to experiment}
\end{figure}

A continuous wavelet transform (CWT) was applied to the pressure signals on the blades and in the draft tube to evaluate the performance of the CFD models in the frequency domain. For the pressure sensors, a comparison of the dominant frequencies is shown in Figure \ref{fig:CWT-runner}. In the figures, the timing is from the simulation start, which is 1.16 s after the start-up is initialized due to the minimum guide vane opening in the simulation. Figures \ref{fig:RB_PS1SST}, \ref{fig:RB_PS1SIM},  \ref{fig:RB_PS1RSM} present the scalogram of the simulated pressure signals at the leading edge RB PS1, while Figure \ref{fig:RB_PS1EXP} shows the corresponding scalogram obtained from the experimental measurements. In these figures, the red dotted lines indicate the rotor-stator interaction (RSI) from the rotating frame at 20$f_n$, as well as the first and second harmonics of the runner rotational frequency at $f_n$ and 2$f_n$. The RSI at 20$f_n$ appears with high intensity in the simulations (Figures \ref{fig:RB_PS1SIM} and \ref{fig:RB_PS1SIM}) and aligns well with the frequency observed in the experimental data as shown in Figure \ref{fig:RB_PS1EXP}. In addition, the first harmonic of the runner rotational frequency is visible in both the simulations and the experiment, reaching approximately 15 Hz. Higher-order harmonics of the rotational frequency, including the second, third, and fourth harmonics, are clearly observed in the experimental signal. In contrast, only the second harmonic is discernible in the simulations, where it appears weak.


A comparison of the pressure signals in the draft tube between the experiment and the simulations is shown in Figures  \ref{fig:p9356_EXP}, \ref{fig:SST_p9356_SIM}, \ref{fig:p9356_SIM} and \ref{fig:RSM_p9356SIM}. In the draft tube, frequency content is observed in the range of 19 to 50 Hz, which may be associated with the presence of multiple vortices and their harmonics. At the times corresponding to the highest pressure amplitudes, the runner rotational frequency ranges between 9 and 13 Hz. In the experiment, high-amplitude pressure fluctuations are detected in the frequency range from 19 to 50 Hz, but with the most dominant frequency of 26 Hz. At the corresponding time, the BSL-RSM simulation shows a similar frequency range with the same constant dominant frequency around 26 Hz. The BSL-EARSM and the k-$\omega$ SST simulations do, in contrast, not show the same range of frequencies but the BSL-EARSM shows a narrow frequency band  which increases from 26 Hz to 30 Hz before fading and the k-$\omega$ SST shows an elevated amplitude at 26 Hz which then drops to 13 Hz shortly after. This close alignment for the BSL-RSM indicates the model's ability to capture the draft tube vortex dynamics. The BSL-EARSM simulation, however, does not manage to capture the complex vortex dynamics present in the draft tube fully. 

The scalograms of the pressure signal at the trailing edge of the runner blade are shown in Figures \ref{fig:RB_PS5EXP}, \ref{fig:RB_PS1SST}, \ref{fig:EARSM_RB_PS5SIM}  and \ref{fig:RSM_RB_PS5SIM}. The first harmonic of the runner rotational frequency is visible in both the experimental and simulated signals, whereas the second, third, and fourth harmonics observed in the measurements are not captured by the simulation. This discrepancy may indicate the presence of fluid-structure interaction effects along the blade that are not reproduced by the numerical model (\cite{biner_fatigue_2024}). In the experiment, at approximately the same time when partial vortex frequencies are observed in the draft tube between 3 s and 4 s, a high-amplitude pressure oscillation at 44 Hz is detected at the trailing edge sensor RB PS5. This frequency aligns well with the fourth harmonic of the runner rotational frequency, which lies between 38 and 44 Hz during this phase of the start-up. The high amplitude observed at this stage may therefore be associated with resonance between the fourth harmonic of the runner rotation and the harmonics of either the vortex ropes in the draft tube or the inter-blade vortices, which are observed in the frequency range from 19 to 50 Hz. However, distinguishing between these contributions is challenging due to the presence of multiple vortical structures and uncertainty in the exact vortex frequencies due to constantly changing rotational speed and flow conditions. A similar resonance behaviour is suggested by a comparison between the draft tube pressure sensor and the trailing edge pressure sensor in the simulations, between Figures \ref{fig:p9356_SIM} and \ref{fig:EARSM_RB_PS5SIM} for the BSL-EARSM simulation and Figures \ref{fig:RSM_p9356SIM} and \ref{fig:RSM_RB_PS5SIM} for the BSL-RSM simulation. Although the BSL-EARSM and BSL-RSM models show this high-amplitude oscillation, the excited frequencies of the BSL-RSM simulation (Figure \ref{fig:RSM_RB_PS5SIM}) match the experiment better in frequency than the BSL-EARSM simulation (Figure \ref{fig:EARSM_RB_PS5SIM}). The k-$\omega$ SST, however, fails to show this behaviour in the draft tube and the blades. 

\section{Conclusion}

In this article, the feasibility of simulating the linear turbine start-up of a variable-speed pump-turbine with URANS in a computationally tractable way is assessed and compared to reduced-scale model experimental data. The simulations, which required 34.560 core hours for the BSL-EARSM and the k-$\omega$ SST and 56.160 core hours for the BSL-RSM, are all able to capture the main flow evolution of the torque, discharge, and leading edge pressure with reasonable accuracy when compared to experimental data, both in time and frequency domains. During the early stages of the start-up, both the k-$\omega$ SST and the Reynolds Stress Model (RSM) exhibited slight numerical oscillations that were not present in the experimental data, with the oscillations being more pronounced in the k–$\omega$ SST case. However, during the high-pressure pulsation regime, the k-$\omega$ SST and the BSL-RSM were able to match the amplitude of the pressure oscillations better at the trailing edge and in the draft-tube. Overall though, the BSL-RSM showed the closest agreement. In the frequency domain, the BSL-RSM showed superiority over the BSL-EARSM and the k-$\omega$ SST, where it managed to match the dominant frequencies at the trailing edge and in the draft tube considerably better, as it managed to capture the range of frequencies during the high-amplitude regime, whereas the other models only resolved the most dominant frequency. Future work will focus on validating the flow fields and the correct development of vortex structures in the draft tube through experimental data to gain a better understanding of model performance and the underlying flow mechanisms leading to runner blade stress increase. 

\section{Acknowledgement}
This work is based on an experiment conducted within the XFLEX Hydro project and has received funding from the European Union’s Horizon Europe research and innovation programme under grant agreement No. 857832. The authors would like to thank Martin Seydoux for conducting the experiment which the CFD models are validated and compared with.

\printbibliography[heading=bibintoc,title={Bibliography},sorting=none]
\end{document}